# Contrast Enhancement of Binary Star System Using an Optical Vortex Coronagraph

*Grover A. Swartzlander, Jr., Erin L. Ford, Rukiah S. Abdul-Malik*
*College of Optical Sciences, University of Arizona, Tucson, Arizona*

*Laird M. Close, Mary Anne Peters*
*Steward Observatory, University of Arizona, Tucson, Arizona*

*David M. Palacios, Daniel W. Wilson*
*Jet Propulsion Laboratory, California Institute of Technology, Pasadena, California*

Send correspondences to: grover.swartzlander@gmail.com

**Abstract:** Using an optical vortex coronagraph and simple adaptive optics techniques we have made the first convincing demonstration of an optical vortex coronagraph that is coupled to a star gazing telescope. In particular we suppressed by 97% the primary star of a barely resolvable binary system, Cor Caroli, having an effective angular separation of only $1.4\lambda/D$. The secondary star suffered no suppression.

## 1. Introduction

Since the early discoveries of planets orbiting distant stars [1,2], and indeed before then [3], astronomers have pondered the possibility, while wrestling with the requirements, of detecting biomarkers from distant worlds [4], e.g., spectral lines of oxygen and water vapor. Current efforts by the U.S. National Aeronautics and Space Administration (the Terrestrial Planet Finder mission) [5] and the European Space Agency (the Darwin mission) [6] to directly detect light from earth-like planets is confounded by the tremendous glare from the parent star. Not only is the planet expected to be roughly $10^7$ times less intense that the starlight in the near-infrared spectrum ($10^{10}$ in the visible), the angular separation is expected to be small in the habitable zone: ~100 mas. One approach to this problem is the development of an ultra-high contrast coronagraph that suppresses starlight while leaving intact the planet light. Various schemes have been proposed, analyzed [7], and tested in the laboratory [8-14]. Reproducing such measurements on a star gazing telescope is challenging, owing in part to atmospheric turbulence, pointing stability, and long integration times. Here we report actual telescope measurements using a promising high contrast imaging instrument, the optical vortex coronagraph.

Similar in design to a Lyot coronagraph [15] an optical vortex coronagraph [16, 17] is able to suppress the glare from a bright point object so that fainter adjacent objects can be better imaged. Whereas a Lyot coronagraph employs an opaque occulting disk to block the central Airy spot from reaching an imaging camera, a vortex coronagraph takes advantage of the diffractive properties of an optical vortex lens [18-20]. Both systems employ a "Lyot stop" to remove undesirable diffracted light from the imaging system. Whereas the performance of the traditional Lyot coronagraph suffers from significant leakage of unwanted light through the Lyot stop, the vortex coronagraph has the potential to completely reject light all the light from a point-like star without diminishing light from objects adjacent to the star. Laboratory measurements have achieved contrast enhancements at $4\lambda/d$ of greater than $10^4$ [21]. Many challenges must be surmounted before such values can be achieved on a star gazing telescope. Most importantly the star must have a stationary aberration-free focus on the center of the vortex lens. In particular we must stop the dancing of the stars, which is caused by atmospheric density fluctuations. We have achieved stationary high Strehl imaging by use of a small aperture Keplarian telescope and a fast external tip-tilt correcting mirror. When the entrance pupil of the telescope is smaller than the Fried parameter, high speed tip-tilt corrections of the wavefront can deliver a Strehl ratio of roughly 99% [22].

In Section 2 we describe the design of the apparatus, and in Section 3 we present our experimental results and analysis.

## 2. Vortex Coronagraph Design

A schematic diagram of an optical vortex coronagraph is shown in Fig. 1. The primary star ($S_1$) is focused on the axis of an optical vortex lens (OVL) whereas the secondary star ($S_2$) is imaged elsewhere on the OVL. Substantial spatial filtering of $S_1$ is achieved by placing an overfilled aperture, i.e., a Lyot stop (LS), in the plane of the exit pupil of the system. Light from this on-axis star is, in principle, completely diffracted outside the exit pupil, whereas the off-axis light sources uniformly fill the exit pupil. The Lyot stop therefore only transmits light from $S_2$ to the imaging camera ($Cam_2$). In practice the moving inhomogeneous atmosphere (Atm) severely distorts the wavefronts at the entrance pupil (or aperture, Ap) of the telescope; hence, an ideal stationary image of $S_1$ does not form at the axis of the OVL. Fortunately, the wavefront at the entrance pupil predominantly suffers only tip-tilt distortion when the radius of the entrance pupil, $R_{ap}$, is less than the Fried parameter, $r_0 \sim 100$ mm. Furthermore, atmospheric fluctuations giving rise to tip-tilt distortion occur on the order of $T_{atm} \sim 0.01$ sec [23]. Using an optical detector and a feed-back system with a response time $\tau \ll T_{atm}$, one may construct an adaptive optics system that achieves near-perfect seeing when $R_{ap} < r_0$ [23].

We have employed such a tip-tiled corrector to prevent the image of $S_1$ from wandering away from the axis of the optical vortex lens. This is accomplished by forming an image of the entrance pupil on a piezoelectrically driven tip-tilt mirror (TTM) which corrects the wavefront by means of a high speed camera ($Cam_1$) [Andor iXon EMCCD] and a feedback signal from the camera to TTM (via LabView software and a computer). A Pellicle beamsplitter reflects roughly half of the beam to the high speed imaging camera and the remainder is transmitted to a long exposure telescope camera [SBIG ST-402ME]. To ensure that that aperture Ap forms the required image at the lens $L_2$ (the entrance of the coronagraph system) we found it necessary to insert an optical relay comprised of lenses $RL_1$ and $RL_2$. The parameters of our optical elements are listed in Table 1.

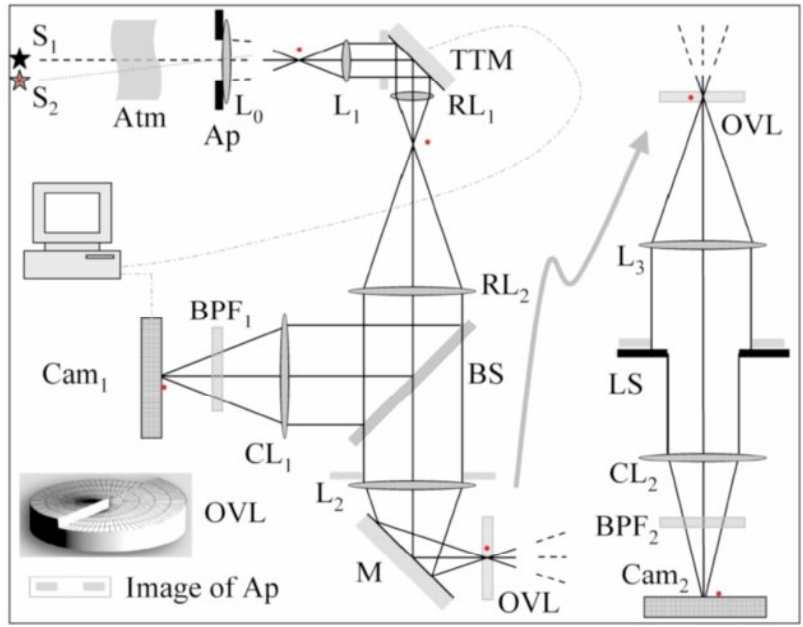

Fig. 1. Schematic diagram showing rays from a binary star system ($S_1$, $S_2$) passing through the atmosphere (Atm) into a telescope comprised of an objective lens, $L_0$, and subaperture, Ap. The rays are recollimated by lens $L_1$ whereby an image of Ap appears at the piezo-electrically driven tip-tilt mirror (TTM). A relay system, $RL_1$ and $RL_2$, reproduces this image at the input face of lens $L_2$. A 50/50 beam splitter (BS) directs light to a fast (500 Hz) camera, $Cam_1$, whereby the star system is imaged by use of camera lens ($CL_1$). The remaining light is imaged on the optical vortex lens (OVL) by use of $L_2$ and recollimated by lens $L_3$. A Lyot stop (LS) is positioned in the image plane of Ap, and the unstopped light forms a spatially filtered image of the star system on a telescope camera ($Cam_2$) by used of the lens $CL_2$. Both cameras employ bandpass filters (BPF). The surface of an OVL is depicted in the inset.

| Element | Ap | $L_0$ | $L_1$ | $RL_1$ | $RL_2$ | $CL_1$ | $L_2$ | OVL | $L_3$ | LS | $CL_2$ |
|---------|------|-------|-------|--------|--------|--------|-------|-----|-------|------|--------|
| D       | 25.4 | 200   | 12.7  | 12.7   | 25.4   | 12.7   | 25.4  | 2.0 | 25.4  | ≤ 3  | 12.7   |
| f       |      | 1.7 m | 50    | 25     | 100    | 50     | 100   |     | 100   |      | 50     |

Table 1. Optical Design Parameters: diameter, D, focal length, f. All lenses except $L_0$ are near-infrared achromatic doublets. Units are millimeters unless stated otherwise.

The 8-inch refractive finder telescope on the Raymond E. White telescope in the Steward Observatory on the campus of the University of Arizona was used as our objective, $L_0$. This telescope system is equipped with excellent tracking and "go to" controls. Further we were able to properly balance the telescope to accommodate our 50 lb system. All other optics had broadband antireflection coatings (including one side of the vortex lens), the surfaces are figured for wavefront errors of $\lambda/10$ or better, and the lenses were achromatic doublets designed for use in the vicinity of 800 nm. The imaging cameras employed bandpass filters, $BPF_1$ and $BPF_2$, designed for $800 \pm 20$ nm. We chose this region of the spectrum owing to a large values of $r_0$ compared to visible wavelengths, and owing to an absence of strong atmospheric absorption lines, hydroxyl fluorescence lines, and thermal background intensity. A narrow band filter was required because the OVL is chromatic [24]. To afford optimal alignments, both the OVL and LS where mounted on separate multi-axis translation stages. The optomechanical platform was analyzed for stiffness using professional software (Cosmos SolidWorks). We calculated a worse case scenario optomechanical error of 0.006 waves at a telescope inclination of 70 deg.

The optical vortex lens was fabricated at the Jet Propulsion Laboratory using analog electron beam lithographic techniques [25]. It is comprised of spin coated PMMA photoresist on a BK7 glass window having a small 5-arcmin wedge angle. One side of the substrate had a broadband near infrared antireflective coating. The surface quality was measured with a phase shifting interferometric profilometer (Veeco NT 9800). The central region near the vortex axis, shown in Fig. 2(a), is relatively smooth. In comparison, the zoomed out image in Fig. 2(b) reveals "stitching errors" which are typical of large area patterns written with an electron beam. Atomic force microscope measurements in the vicinity of the vortex core indicate that the surface pitch is $\Delta d = 1.661$ μm. Assuming the refractive index of PMMA is 1.486, the "design" wavelength $\lambda_0$ at which the doubly pitched surface produces a vortex of topological charge $m(\lambda_0) = 2$ is $\lambda_0 = (\Delta d)(n_{PPMA}-1) = 807$ nm [17]. The topological charge at other wavelengths is given by $m = m_0 * \lambda_0 / \lambda$. At the wavelength corresponding to the center of the bandpass filter the topological charge is therefore expected to be $m(\lambda = 800 \pm 20$ nm$) = 2.02 \pm 0.05$.

The surface roughness over the central patch of the vortex lens may be characterized by the standard deviation in the height, $\sigma_h$. Using a Veeco profilometer we determined that $\sigma_h$ is equivalent to $5/1000^{th}$ of a wave (at the design wavelength). This error is small in comparison to the 3.5% topological charge error owing to the bandwidth and mismatch of our spectral filter. However, the lithographic stitching errors produced undesirable surface deviations of 2 to 5%. The rectilinear pattern of the latter may diffract unwanted light from the on-axis star through the Lyot stop and therefore degrade the device performance.

The rectangular dimensions of the vortex lens is 2 mm x 2 mm. According to the values in Table 1 the re-imaged pupil at lens $L_2$ has a diameter of $D_2 = 3.0$ mm. Hence the diameter of the focal spot on the vortex lens has a calculated value of $2 \times 1.22 \lambda f_2 / D_2 = 65$ $\mu m$. As evident from an examination of Fig. 2(a), this focal spot falls within a single e-beam writing patch on the fabricated vortex lens. Any light from the primary star that is scattered from the e-beam stitching errors can therefore be attributed to light from the Airy rings, not from the central Airy disk. The area of the vortex lens therefore accommodates roughly (1 mm)/$(\lambda f_2 / D_2) = 37.5$ Airy rings across its surface. The secondary star lays 19.3 arcsec from the primary, but the relay system magnifies this by a factor of four. At the vortex lens the primary and secondary are therefore separated by 77.2 arcsec, which is equivalent to $1.40 \lambda / D_2$ where $\lambda / D_2 = 267$ $\mu$rad $= 55$ arcsec. This angular separation is equivalent to 37.4 μm on the vortex lens. Therefore the image of the secondary star on the vortex lens is expected to appear near the first minimum of the Airy ring of the primary star. That is, the two objects are slightly beyond the resolution criterion. The close proximity of these two bodies constitutes a small inner working angle (compared to say $4\lambda / D_2$) and is an exceptionally challenging regime for any high contrast coronagraph.

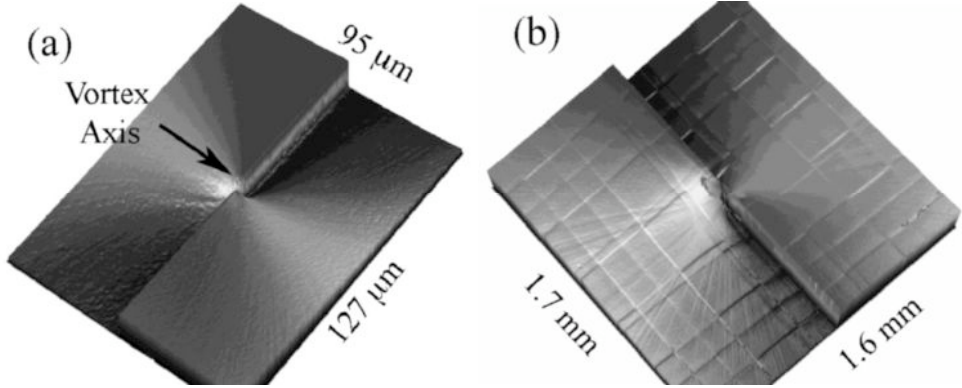

Fig. 2. Surface profile of the optical vortex lens in photoresist showing (a) relatively good surface quality in the central region containing the vortex axis, and (b) significant surface roughness owing to stitching errors across element.

## 3. Experimental Results

Early experiments on a large aperture ($R_{ap}$ = 60 inch) demonstrated the need to minimize the effects of poor atmospheric seeing [26]. In most camera exposures the star focused well away from the vortex axis. A rare case where the two coincided is shown in Fig. 3(a). Clearly there is a need to apply for adaptive optics to our system. For comparison an image of a star using our high Strehl vortex coronagraph is shown in Fig. 3(b).

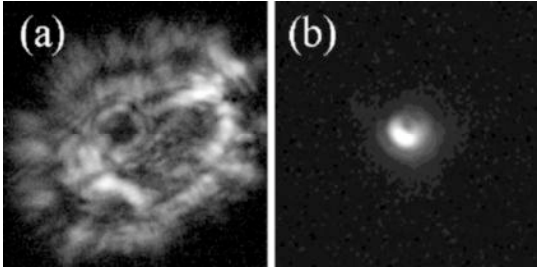

Fig. 3. (a) Image of a star (Betelgeuse) through a vortex lens without adaptive optics [25]. (b) Image of a star (Arcturus) through a vortex lens using adaptive optics.

Before testing the nulling capacity of the coronagraph on a binary star system we tested the system in the laboratory using a fiber-coupled diode laser as a light source. Without the vortex lens in place we obtained a high Strehl image showing many Airy rings, as seen in Fig. 4(a). With the vortex lens in place the central Airy disk assumed a ring profile, as shown in Fig. 4(b). Relative to the peak of the Airy disk, the vortex lens suppressed the center peak by roughly a factor of 1/400, and the average maximum of the ring was 1/200 times smaller. More importantly, the contrast at the fourth Airy ring corresponds to a contrast of $2.3 \times 10^4$, as is evident by the azimuthally averaged plots in Fig. 4(c).

To test the nulling capability of the OVC when it is coupled to a star gazing telescope, we pointed the telescope at the binary star system, Cor Caroli (alf 2 CVn) in the constellation Canes Venatici. The stars have a known separation of 19.3 arcsec and have a visible magnitude difference of $\Delta M$=2.7 (i.e., $100^{(2.71/5)}$ = 12.1 times flux ratio), with the primary star having a magnitude of 5.6. It is an infrared source and is 110 light years from Earth. The primary star is a variable star with visual amplitude change of up to 0.1 [27]. The primary is labeled $a^2$ or alpha$^2$ since it is slightly east of its companion $a^1$ or alpha$^1$ [28].

We recorded images and dark frames at 300 second exposures through an 800 nm ± 20 nm bandpass spectral filter. When the vortex lens was misaligned the strong primary ($a^2$) and the weak secondary ($a^1$) stars can be resolved in Fig. 5(a). This false color image is unaltered data from which a dark frame image has been subtracted. The absence of Airy rings around the stars suggests that our tip-tilt corrector was unable to correct all the perturbations. This is understandable because the telescope was looking through a thick atmosphere (~30 deg above the horizon) and the telescope was not well balanced at that angle. Alternatively the image image may have been slightly defocused. The later is supported in part by our numerical simulations (not shown).

After aligning the center of the vortex lens with the primary star the relative magnitudes of both stars were rendered nearly equal, as can be seen in Fig. 5(b). The difference between these Fig. 5(a) and (b), shown in Fig. 1(c), suggests that the signal from from the secondary star is nearly unaffected by the vortex lens, whereas the primary star experiences a significant reduction.

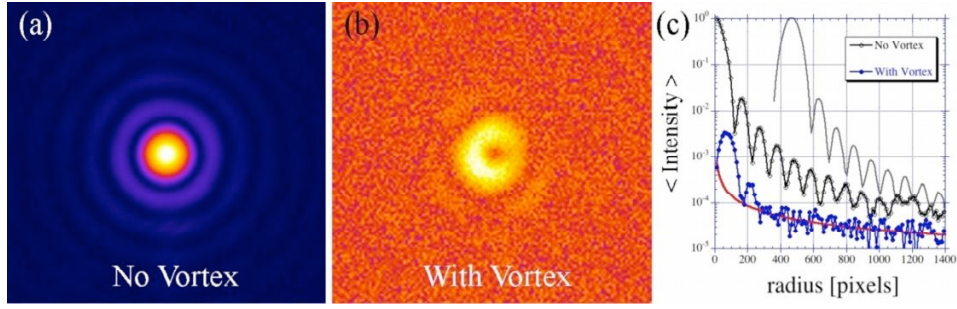

Fig. 4. False color laboratory measurements using a vortex coronagraph. (a) Without vortex lens. (b) With vortex lens. (c) Radially averaged intensity profiles of the data in (a) and (b).

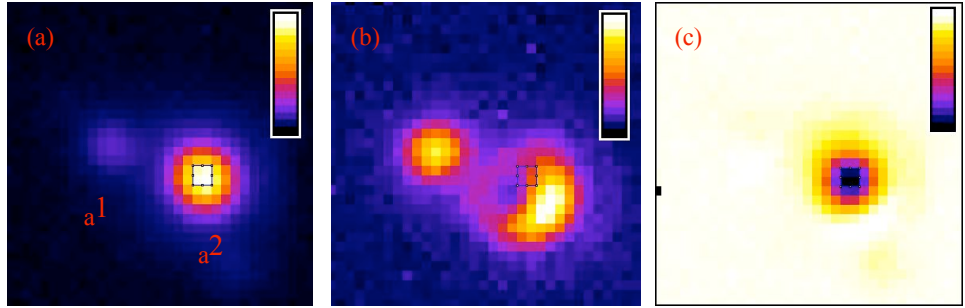

Fig. 5. Raw dark frame subtracted images in false color (a) without vortex, (b) with vortex, (c) difference with (a) subtracted from (b). The 2x2 pixel outline in each image demarks the location where the difference is greatest.

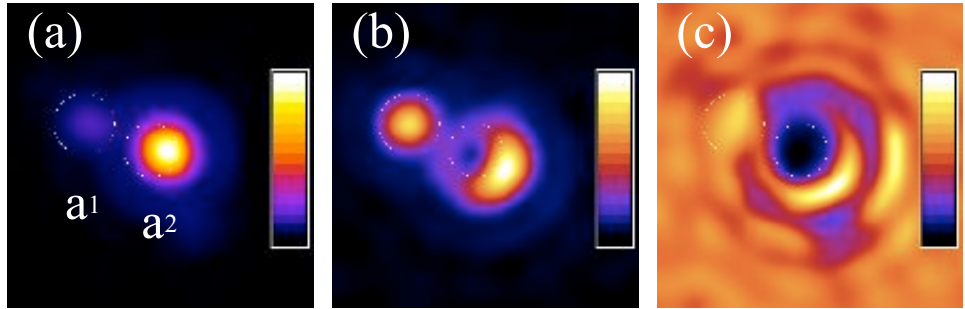

Fig. 6. Interpolated images (a) without vortex ($f_a$), (b) with vortex lens ($f_b$), (c) Relative difference, $(f_b-f_a)/f_a$.

Using a Fourier interpolation technique (zero padding in the transformed domain) we can obtain a clearer impression of the recorded images, as shown in Fig. 6. This interpolation affords an ability to make calculations and line plots with less pixelation noise. High spatial frequency noise has been numerically filtered from the images by use of a supergaussian soft aperture. These images allowed us to plot the intensity profiles shown in Fig. 7. The black (blue) line corresponds to the intensity without (with) the vortex aligned with the primary star. Each of the plots corresponds to the same straight line in each image. This line of data was selected by tracing a line segment through the peaks in Fig. 6(a). Without the vortex lens the "uncorrected" intensity peak of the primary (secondary) star has a value of 467.57 (73.39) in arbitrary units. The dark floor for this case was found to be 9.75±0.41. We note that the ratio of the corrected peaks (measured at 800 nm) is equal to 457.8/63.6 = 7.2 (or ΔM=2.14), which is less than the reported value in the visible spectrum (12.1).

After introducing the vortex lens the intensity peak of the primary star is spatially shifted. The uncorrected peak value of the primary

(secondary) star is 84.20 (73.73), with a dark floor of 7.67±0.32. Here we are concerned with the maximum suppression afforded by the vortex lens. We therefore subtract the dark floor from both signals and calculate the point-by-point percent change. We find the maximum percent change is 97%, as indicated by the red line in Fig. 7.

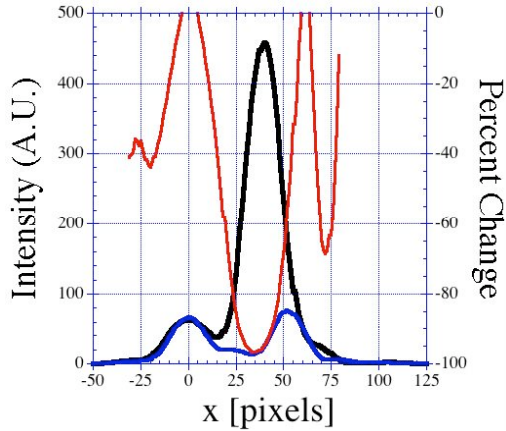

Fig. 7. Line plots showing (black line) the peaks of the primary and secondary star, (blue line) the suppression owing to the vortex lens, and (red line) the percent change that reaches -97%.

**4. Conclusion**

We have constructed and mounted to a telescope an optical vortex coronagraph coupled to a tip-tilt correcting adaptive optics system. The system was designed to produce a Strehl ratio of ~99% and to maintain a stationary focal spot on the axis of an optical vortex lens. Images of a binary star system with an effective angular separation of 1.4λ/D were recorded with and without the vortex lens in place. The intensity of the primary star was suppressed by 97% whereas the intensity of the secondary star was unaffected. In comparison the system achieved a suppression of 99.8% in the laboratory. We attribute the difference to systematic and human errors, e.g., poor seeing through thick early morning atmosphere (~30 deg above the horizon) and misaligned optics. Future work requires the fabrication of higher quality vortex lenses, improved optomechanical systems, and the inclusion of advanced adaptive optics techniques. A larger telescope aperture or lower noise cameras would help us avoid dark noise limitations. We also suggest viewing systems that are more readily resolvable (e.g., ~4λ/D) and avoid systems that are close to the horizon.

**Acknowledgements**
We gratefully acknowledge assistance from Joshua Nelson, Tom Fleming, Dave Harvey, Chris Limbach, Gary Rosenbaum, and Roger Ceragioli (Steward Observatory, University of Arizona). We are also grateful to Stewart B. Shaklan and Wesley Traub (Jet Propulsion Laboratory, California Institute of Technology), and Olivier Guyon (Subaru Telescope, National Astronomical Observatory of Japan). We are thankful to Joanna Schmit (Veeco Inc., Metrology Group, Tucson, AZ) for profilometric measurements. We also acknowledge earlier vortex coronagraph collaborations with Gregory E. Foo (Northrop Grumman Space Technology, Long Beach, CA) and Eric Christensen (Lunar and Planetary Laboratory, University of Arizona) using the 60" telescope on Mt. Lemmon, Arizona, in 2002.

The U.S. Army Research Office and the Jet Propulsion Laboratory supported the effort in the College of Optical Sciences. The Steward Observatory effort was supported by NASA Origins and the National Science Foundation Career program. The Jet Propulsion Laboratory effort was supported under contract with the National Aeronautics and Space Administration (NASA).

**References**
[1] A. Wolszczan, D. A. Frail, "A planetary system around the millisecond pulsar PSR1257+12," Nature 355, 145-147 (1992).
[2] G. W. Marcy and R. P. Butler, "Detection of extrasolar giant planets," Annu. Rev. Astron. Astrophys. 36, 57-97 (1998).
[3] S. J. Dick 1996 *The Biological Universe* (Cambridge Univ. Press, Cambridge, 1996).
[4] V. G. Ford, P. D. Lisman, S. B. Shaklan, J. T. Trauger, T. Ho, D. Hoppe, A. E. Lowman, "The Terrestrial Planet Finder coronagraph: technology and mission design studies," Proc. SPIE 5487, 1274-1283 (2004).
[5] D. Coulter, "NASA/s Terrestrial Planet Finder mission: the search for habitable planets," in Towards Other Earths: DARWIN/TPF and the Search for Extrasolar Planets, M. Fridlund and T. Henning, eds. (European Space Agency, 2003), pp. 47-54.
[6] A. Karlsson and L. Kaltenegger, "The technology of DARWIN," in Towards Other Earths: DARWIN/TPF and the Search for Extrasolar Planets, M. Fridlund and T. Henning, eds. (European Space Agency, 2003), pp. 41-46.
[7] O. Guyon, E. A. Pluzhnik, M. J. Kuchner, B. Collins, S. T. Ridgway, "Theoretical limits on extrasolar terrestrial planet detection with

coronagraphs," Astrophys. J. Suppl. 167, 81–99 (2006).
[8] O. Guyon, C. Roddier, J. E. Graves, F. Roddier, S. Cuevas, C. Espejo, S. Gonzalez, A. Martinez, G. Bisiacchi, and V. Vuntesmeri, "The Nulling Stellar Coronagraph : Laboratory Tests and Performance Evaluation," PASP 111, 1321-1330 (1999).
[9] R. Galicher, O. Guyon, M. Otsubo, H. Suto, and S. Ridgway, "Laboratory Demonstration and Numerical Simulations of the Phase-Induced Amplitude Apodization," PASP 117, 411–420 (2005).
[10] A. Chakraborty, L. Thompson, and M. Rogosky, "$10^{-7}$ contrast ratio at $4.5\lambda/D$: New results obtained in laboratory experiments using nano-fabricated coronagraph and multi-Gaussian shaped pupil masks," Opt. Express 13, 2394-2402 (2005).
[11] D. Mawet, P. Riaud, J. Baudrand, P. Baudoz, A. Boccaletti, O. Dupuis and D. Rouan, "The four-quadrant phase-mask coronagraph: white light laboratory results with an achromatic device," A&A 448, 801-808 (2006).
[12] J. H. Lee, G. Foo, E. G. Johnson, and G. A. Swartzlander, "Experimental Verification of an Optical Vortex Coronagraph," Phys. Rev. Lett. 97, 053901(1-4) (2006).
[13] J. T. Trauger and W. A. Traub, "A laboratory demonstration of the capability to image an Earth-like extrasolar planet," Nature 446, 771-773 (2007).
[14] K. Enya, L. Abe, S. Tanaka, T. Nakagawa, K. Haze, T. Sato, and T. Wakayama, "High contrast experiment of an AO-free coronagraph with a checkerboard pupil mask," Astronomy & Astrophysics, 480, 899-903 (2008).
[15] M. B. Lyot, "A study of the solar corona and prominences without eclipses," Mon. Not. R. Astron. Soc. 99, 580 (1939).
[16] D. Mawet, P. Riaud, O. Absil, and J. Surdej, "Annular Groove Phase Mask Coronagraph," The Astrophysical Journal 633, 1191 (2005).
[17] G. Foo, D. M. Palacios, G. A. Swartzlander, Jr., "Optical Vortex Coronagraph," Opt. Lett. 30, 3308-3310 (2005).
[18] S.N. Khonina, V.V. Kotlyar, M.V. Shinkaryev, V.A. Soifer, and G.V. Uspleniev, "The Phase Rotor Filter (PRF)," J. Modern Opt. 39, 1147-1154 (1992).
[19] G. A. Swartzlander, Jr., "Peering into darkness with a vortex spatial filter," Opt. Lett. 26, 497-499 (2001).
[20] G. A. Swartzlander, Jr., "Broadband Nulling of a Vortex Phase Mask," Opt. Lett. 30, 2876-2878 (2005).
[21] G. A. Swartzlander, Jr., E. L. Ford, R. Abdul-Malik, J. Kim, L. Close, M. A. Peters, D, Palacios and D. Wilson, "Advancements of the optical vortex coronagraph," Proc. SPIE 6693 (Techniques and Instrumentation for Detection of Exoplanets III, Daniel R. Coulter, Editor), 669311-17 (2007).
[22] M. A. Peters, L. M. Close, M. Rademacher, T. Stalcup, G. A. Swartzlander, E. Ford, and R. S. Abdul-Malik, "A high-Strehl low-resolution optical imager (BESSEL): Detection of a $0.7\lambda/D$ separation binary from the ground," New Astronomy 13 359–369 (2008)
[23] Brian Kern, Ted A. Laurence, Chris Martin, and Paul E. Dimotakis, "Temporal coherence of individual turbulent patterns in atmospheric seeing," Appl. Opt. 39, 4879-4885 (2000).
[24]. G. A. Swartzlander, Jr. "Achromatic optical vortex lens" Opt. Lett. 31, 2042-2044 (2006).
[25] D. W. Wilson, R. E. Muller, P. M. Echternach, "Electron-beam lithography for micro and nano-optical applications," SPIE Photonics West, Micromachining Technology for Microoptics and Nanooptics, San Jose, California, January 25-27, 2005 (NASA Technical Report 20060044354).
[25] We acknowledge Gregory E. Foo, Eric Christensen, and Grover A. Swartzlander, Jr. for this measurement using the 60" telescope on Mt. Lemmon Arizona, Dec. 2002.
[27] x.astrogeek.org/observations/log.php?object_id= 508
[28] www.dibonsmith.com/ cvn_con.htm